\documentclass[referee]{aa}
\usepackage{float,graphicx,textcomp}
\usepackage{natbib,amsmath,amssymb}
\usepackage{txfonts}
\usepackage[utf8]{inputenc}
\usepackage{color}
\usepackage{epsfig}
\usepackage{bbm}

\providecommand{\tabularnewline}{\\}

\begin{document}

\title{An advective solar-type dynamo without\\ the $\alpha$ effect}
\titlerunning{An advective solar-type dynamo without the $\alpha$ effect}

\author{N. Seehafer \inst{1} \and V. V. Pipin \inst{2}}

\institute{
Institut f\"ur Physik und Astronomie, Universit\"at Potsdam,
	     Karl-Liebknecht-Str. 24/25, 14476 Potsdam, Germany\\
             \email{seehafer@uni-potsdam.de}
\and
Institute for Solar-Terrestrial Physics,
Siberian Division of the Russian Academy of Sciences, 664033 Irkutsk,
Russia\\
\email{pip@iszf.irk.ru} }

\date{Received .......... /Accepted ..........}

\abstract
{
Most solar and stellar dynamo models use the
$\alpha\Omega$ scenario where the magnetic field is generated by the interplay between differential rotation
(the $\Omega$ effect) and a mean electromotive force due to helical turbulent
convection flows (the $\alpha$ effect). There are, however,
turbulent dynamo mechnisms that may complement the $\alpha$ effect or may be an alternative to it.
}
{
We investigate models of solar-type dynamos where the $\alpha$ effect is completely replaced
by two other turbulent dynamo mechanisms, namely the $\vec{\Omega}\times\vec{J}$ effect and the shear-current effect, which both result from an inhomogeneity of the
mean magnetic field.}
{
We studied axisymmetric mean-field dynamo models containing differential rotation,
the $\vec{\Omega}\times\vec{J}$ and shear-current effects, and a meridional
circulation.
The model calculations were carried out using the rotation profile of the Sun as
obtained from helioseismic measurements and radial profiles of other quantities
according to a standard model of the solar interior.
}
{
Without meridional flow, no satisfactory agreement of the models with the solar
observations can be obtained.
With a sufficiently strong meridional circulation included, however,
the main
properties of the large-scale solar magnetic field, namely, its oscillatory behavior,
its latitudinal drift
towards the equator within each half cycle,  and its dipolar
parity with respect to the equatorial plane, are correctly reproduced.
}
{
We have thereby constructed the first mean-field models of solar-type dynamos that
do not use the $\alpha$ effect.
}


\maketitle

\section{Introduction}

The standard dynamo model for the Sun and stars is the $\alpha\Omega$ model
where, within the framework of mean-field magnetohydrodynamics, the magnetic field
is produced by an interplay between differential rotation (the $\Omega$ effect) and the
collective action of turbulent cyclonic convection flows, known as the $\alpha$ effect \citep{par55,par79,stk66,krarad80}.
The $\alpha$ effect is here responsible for generating
the poloidal component of the large-scale magnetic field (LSMF),
whose toroidal component is mainly generated by the the $\Omega$ effect.
The model is often supplemented with meridional flows, leading
to so-called flux-transport dynamos 
\citep[e.g.,][]{choschusdik95,kukrudschul01,rem06,dikgil07}. The meridional flows may transport
toroidal magnetic flux toward the equator
and their speed may determine the cycle period,
thus allowing us to bypass a number of problems
connected with the $\alpha$ effect
and $\alpha\Omega$ dynamos,
as, for instance, that, in the case of the Sun, the obtained
cycle periods are generally too short and  the magnetic activity is not sufficiently concentrated at low latitudes
\citep[see, e.g.,][]{oss03,rudhol04,brasub05}.

In mean-field magnetohydrodynamics,
the influence of the turbulence on
the LSMF is expressed by the
mean turbulent electromotive force (MEMF), $\vec{\mathcal{E}}=\left\langle \vec{u}\times\vec{b}\right\rangle $,
where
$\vec{u}$ and $\vec{b}$ are the fluctuating parts of the velocity and magnetic
field and angular brackets denote averages.
The by far best known contribution to $\vec{\mathcal{E}}$ is provided
by the $\alpha$ effect, namely,
a turbulent electromotive force $\alpha\langle\vec{B}\rangle$, with $\alpha$ denoting a {(symmetric)} tensorial factor of proportionality and  $\langle\vec{B}\rangle$ the LSMF.
However, there are other turbulent dynamo mechanisms besides the $\alpha$ effect.
Two of them are the $\vec{\Omega}\times\vec{J}$ effect \citep{rad69,sti76}
and the shear-current or $\vec{W}\times\vec{J}$ effect \citep{rogkle03,rogkle04};
$\vec{\Omega}$ is here the angular velocity of the stellar rotation,
$\vec{J}=\nabla\times\langle\vec{B}\rangle/\mu_0$ the large-scale electric-current density, and $\vec{W}=\nabla\times\vec{V}$ the
large-scale vorticity, $\vec{V}$ denoting the large-scale velocity.
Both these effects result from an inhomogeneity of the LSMF, in contrast to the $\alpha$ effect, which
also works with a homogeneous $\langle\vec{B}\rangle$
(that is to say, for calculating the $\alpha$ effect, $\langle\vec{B}\rangle$ may be considered as homogeneous on the scale of the fluctuations).

In the commonly used representation of the MEMF
on the basis of symmetry considerations
 \citep[see][]{rad80,krarad80,rad00,radklerog03},
 the $\vec{\Omega}\times\vec{J}$ and shear-current effects represent contributions to the $\vec{\delta}$ term,
{a term of the form
$\vec{\mathcal{E}}_\delta=\vec{\delta}\times(\nabla\times\langle\vec{B}\rangle)$, where $\vec{\delta}$ is a vector. Since $\vec{\mathcal{E}}_\delta\cdot\vec{J}=0$,
the effects described by this term cannot bring energy into the mean magnetic field
and, thus, cannot lead to working dynamos when acting alone. }
These effects have been investigated little in the context of solar and stellar dynamos.
For a recent study of the possible role of the $\vec{\Omega}\times\vec{J}$ effect when acting together with the $\alpha$ effect and differential rotation in a  spherical shell, {or when acting together with another part of the MEMF, not included in
dynamo studies before,}
in a rigidly rotation full sphere, see \citet{pipsee09}, where an illustration of the physical mechanism behind the $\vec{\Omega}\times\vec{J}$ effect also may be found; the mechanism of the
shear-current effect is very similar to that of the $\vec{\Omega}\times\vec{J}$ effect.

In this paper, we consider mean-field dynamo models in the geometry
of a spherical shell, as appropriate for solar-type stars, where the $\alpha$ effect
is completely omitted. Instead, the $\vec{\Omega}\times\vec{J}$ and shear-current effects serve as turbulent dynamo mechanisms.
In nearly all mean-field dynamo studies, the effective strengths of the different physical ingredients are controlled by freely varied dimensionless parameters; in the case
of dynamo effects, e.g., the $\alpha$ effect, these are usually referred to as
dynamo numbers.
This reflects our present knowledge of the physical processes in the convection
zones of the Sun and stars. Realistic self-consistent numerical models of these processes and their interactions will remain out of reach for the foreseeable future.
Given this situation, we deem it advisable to explore the potentials of
turbulent dynamo effects other than the $\alpha$ effect.

{
Numerical evidence for turbulent dynamo effects has so far mainly been obtained from convection
simulations in small (compared to the dimensions of a star) rectangular boxes
\citep[e.g., ][]{braetal90,ossstibra01,ossetal02,giezierud05,kapetal06,cathug06,hugcat08}.
Due to the assumption of a uniform mean magnetic field and other limitations, most of
these studies could only find parts of the MEMF that are proportional to the LSMF,
i.e., the $\alpha$ effect and turbulent pumping
(a contribution to the MEMF of the form
$\vec{\mathcal{E}}_\gamma=\vec{\gamma}\times\langle\vec{B}\rangle$,
with $\vec{\gamma}$ denoting a vector;
it leads to an advection of the mean magnetic field).
Recently, however, \citet{kapkorbra09}, who used a procedure referred to as the test field
method \citep{schrietal05,schrietal07} together with numerical simulations of turbulent convection with shear
and rotation, were able to also identify the action of the combined $\vec{\Omega}\times\vec{J}$ and shear-current effects.
}

Here, we explore axisymmetric kinematic dynamo models containing the
$\vec{\Omega}\times\vec{J}$ and shear-current effects, differential rotation,
and meridional circulation. In calculating the MEMF we use analytical expressions 
derived by \cite{pip08}
on the basis of a simplified version
of the $\tau$ approximation
{\citep[cf.][]{vaikic83,brasub05,brasub05b}}.
We construct models with distributed dynamo action in the
bulk of the convection zone, rather than in the overshoot layer at the bottom of the convection zone.
The model calculations are carried out using the rotation profile of the Sun as
obtained from helioseismic measurements and radial profiles of other quantities
according to a standard model of the solar interior.

The remainder of the paper is organized as follows: In Sect.~\ref{sec_model} we describe our dynamo model,
as well as the used numerical procedure
(some benchmark tests for our computer code are presented in Appendix \ref{appendix_benchmarks}).
 Then, in Sect.~\ref{sec_results}, we present the obtained results.
In Sect.~\ref{sec_conclusions}, we draw conclusions and discuss our results.

\section{Model and numerical procedure}
\label{sec_model}

The axisymmetric LSMF is written in the usual way as the sum of a poloidal and a toroidal part,
\begin{equation}
\left\langle \vec{B}\right\rangle =\nabla\times \frac{A\vec{e}_{\phi}}{r\sin\theta} +B\,\vec{e}_{\phi}\,,
\end{equation}
where $A(r,\theta,t)$ (the flux function for the poloidal field) and 
$B(r,\theta,t)$ (the toroidal field component) are scalar functions of radius,
$r$, colatitude, $\theta$, and time, $t$, and  $\vec{e}_{\phi}$ is
the unit vector in the direction of the azimuthal coordinate, $\phi$.
The mean-field induction equation then takes the form
\begin{align}
\frac{\partial A}{\partial t} & =  r\sin\theta\,\mathcal{E}_{\phi}-\frac{U_{\theta}}{r}\frac{\partial A}{\partial\theta}-U_{r}\frac{\partial A}{\partial r}\,,\label{eq:2}\\
\frac{\partial B}{\partial t} & =  \frac{1}{r}\frac{\partial\left(\Omega,A\right)}{\partial\left(r,\theta\right)}+\frac{1}{r}\left(\frac{\partial r\left(\mathcal{E}_{\theta}-U_{r}B\right)}{\partial r}-\frac{\partial\left(\mathcal{E}_{r}+U_{\theta}B\right)}{\partial\theta}\right)\,,\label{eq:1}
\end{align}
where the effects of the large-scale flows enter
through the differential-rotation
rate, $\Omega(r,\theta)=|\vec{\Omega}(r,\theta)|$, and the components
  of the meridonal flow, $U_{r}$ and $U_{\theta}$.

To calculate the MEMF, whose effects appear through the components
of $\vec{\mathcal{E}}$ in Eqs. (\ref{eq:2}) and (\ref{eq:1}), we modify the expressions given in \citet{pipsee09} by completely omitting the $\alpha$ effect
but including, in addition to the $\vec{\Omega}\times\vec{J}$ effect, isotropic
and anisotropic turbulent diffusion, and turbulent pumping,
now also the
shear-current effect.
The contribution of this
to the MEMF is to linear order, i.e., for a weak mean magnetic field, as well
as neglecting the effect of the Coriolis force,
given by \citep{pip08}
\begin{equation}\label{difs1}
\begin{split}
\mathcal{E}_{i}^{(W)}=\varepsilon_{inm}\Big\{& C_{1}\overline{V}_{lm}\overline{B}_{nl}+C_{2}\overline{B}_{nl}\overline{V}_{ml}+C_{3}\overline{V}_{lm}\overline{B}_{ln} \\
 &+C_{4}\overline{B}_{ln}\overline{V}_{ml}\Big\} \left\langle {u^{(0)}}^2\right\rangle \tau_{c}^{2} \,,
\end{split}
\end{equation}
where tensor notation and the summation convention have been used.
$\overline{B}_{ij}=\partial\langle B_i\rangle/\partial x_j$ is the gradient tensor
of the mean magnetic field and $\overline{V}_{ij}$ a corresponding quantity
for the differential rotation, namely,
 $\overline{V}_{ij}=\partial V_i/\partial x_j$,
with $\vec{V}=r\sin\theta\left(\Omega-\Omega_{0}\right)\vec{e}_{\phi}$
being the rotational velocity, defined
in a reference frame that rotates with
angular velocity $\Omega_{0}$, the rigid-body rotation rate of the core
(encountered at midlatitudes through the convection zone).
$\vec{u}^{(0)}$ is a small-scale or turbulent convective background velocity as present
in the absence of rotation and a mean magnetic field,
$\tau_{c}$ the correlation time of $\vec{u}^{(0)}$,
and
 $C_{1}=\left(\varepsilon-3/5\right)/6$, $C_{2}=\left(\varepsilon-1\right)/5$,
$C_{3}=\left(1+\varepsilon\right)/15$,
and $C_{4}=-\left(7\varepsilon+11\right)/30$ are constants;
$\varepsilon=\sqrt{\left\langle {\vec{b}^{(0)}}^{2}\right\rangle }/\left(u_{c}\sqrt{\mu_0\rho}\right)$ is
the square root of the prescribed ratio between the energies
of a fluctuating magnetic background field $\vec{b}^{(0)}$, assumed to be generated by a
small-scale dynamo (which is fully independent of the mean magnetic field), and the background velocity field $\vec{u}^{(0)}$
($u_c=\sqrt{\left\langle{\vec{u}^{(0)}}^2\right\rangle}$ is the rms value
of the convective background velocity field and $\rho$ the mass density).

In the following, we assume energy equipartition between the two background
fields, i.e., $\varepsilon=1$.
Furthermore,
only the azimuthal
$\vec{\Omega}\times\vec{J}$ and shear-current effects are taken into account. This may be justified by the fact that
the toroidal part of the solar LSMF is much stronger than the poloidal one. However,
the remaining parts of the MEMF (isotropic and anisotropic turbulent diffusion, turbulent pumping) are included in all components.
The components of the MEMF in spherical coordinates then become
\begin{equation}\label{eq:er}
\begin{split} 
\mathcal{E}_{r} = -\tilde{\eta}_{T}&\left\{\frac{f_{2}^{(d)}+2f_{1}^{(a)}\sin^{2}\theta}{r\sin\theta}\frac{\partial\sin\theta B}{\partial\theta}-\frac{f_{1}^{(a)}\sin2\theta}{r}\frac{\partial rB}{\partial r}\right.\\
&\quad+  G\sin2\theta\, f_{1}^{(a)}\, B \,\Biggr\} \,,
\end{split}
\end{equation}
\begin{equation}\label{eq:et}
\begin{split}
\mathcal{E}_{\theta} = \tilde{\eta}_{T}&\left\{\frac{f_{2}^{(d)}+2f_{1}^{(a)}\cos^{2}\theta}{r}\frac{\partial rB}{\partial r}-\frac{2f_{1}^{(a)}\cos\theta}{r}\frac{\partial\sin\theta B}{\partial\theta}\right.\\
 &\quad - G\left(f_{3}^{(a)}+\cos2\theta f_{1}^{(a)}\right)B \Biggr\} \,,
\end{split}
\end{equation}
\begin{equation}\label{eq:ephi}
\begin{split}
\mathcal{E}_{\phi} = \tilde{\eta}_{T}&\left\{\frac{f_{2}^{(d)}+2f_{1}^{(a)}}{r}
\left(\frac{1}{\sin\theta}\frac{\partial^{2}A}{\partial r^{2}}
 +\frac{1}{r^2}\frac{\partial}{\partial\theta}\frac{1}{\sin\theta}\frac{\partial A}{\partial\theta}
\right)\right.\\
 &\quad +G\left(\frac{2\cos\theta f_{1}^{(a)}}{r^2}\frac{\partial A}{\partial\theta}-\frac{f_{1}^{(a)}\cos^{2}\theta+f_{3}^{(a)}}{r\sin\theta}\frac{\partial A}{\partial r}\right)\\
&\left.\quad  + \; C_\delta^{(\Omega)}f_{4}^{(d)}\left(\cos\theta\frac{\partial B}{\partial r}-\frac{\sin\theta}{r}\frac{\partial B}{\partial\theta}\right)\right\} \\
+&\;\mathcal{E}_{\phi}^{(W)} \,,
\end{split}
\end{equation}
with $\mathcal{E}_{\phi}^{(W)}$ in Eq.~(\ref{eq:ephi}) denoting the contribution of the
shear-current effect, given by
\begin{equation}\label{eq:sh-cr}
\begin{split}
\mathcal{E}_{\phi}^{(W)}=
\tilde{\eta}_{T}C_\delta^{(W)} f_{4}^{(d)}
&\left\{\frac{11}{6}\left(\hat{\Omega}-1\right)\left(\cos\theta\frac{\partial B}{\partial r}-\frac{\sin\theta}{r}\frac{\partial B}{\partial\theta}\right)\right.\\
 &\quad + \frac{1}{3}\sin\theta\frac{\partial\left(B,\hat{\Omega}\right)}{\partial\left(r,\theta\right)} \\
 &\quad + \left. \frac{B}{6}\left(\cos\theta\frac{\partial\hat{\Omega}}{\partial r}-\frac{\sin\theta}{r}\frac{\partial\hat{\Omega}}{\partial\theta}\right)\right\} \,.
\end{split}
\end{equation}
$f_1^{(a)}$, $f_3^{(a)}$, $f_2^{(d)}$, and  $f_4^{(d)}$ denote functions of $\varepsilon$ and the Coriolis number $\Omega^{\ast}=2\Omega_0\tau_{c}$
that are given in Appendix \ref{appendix_definitions},
$G=(\partial/\partial r)\log\rho$  is the density scale factor,
$\hat{\Omega}=\Omega/\Omega_{0}$,
 and
 $\tilde{\eta}_{T}=C_{\eta}\,\eta_T$, with
$\eta_T=\left\langle {u^{(0)}}^2\right\rangle \tau_{c}$.
$C_{\eta}\leq1$,
$C_\delta^{(\Omega)}\leq1$, $C_\delta^{(W)}\leq1$
are parameters to control the relative strengths of different turbulence effects.
$C_{\eta}$ regulates the turbulence level,
and
$C_{\delta}^{(\Omega)}$ and $C_\delta^{(W)}$ weight the
 $\vec{\Omega}\times\vec{J}$ effect and the shear-current effect, respectively.

Currently,
the dependence of the shear-current effect on the Coriolis number
is unknown. Eq.~(\ref{difs1}) has been derived disregarding the effect
of the Coriolis force
and is, thus, safely applicable only in the limit of slow rotation,
$\Omega^{*}\ll1$. But in the solar convection zone, in particular its
deeper layers, the Coriolis number is large,
$\Omega^{*}\gg1$ \citep[cf., e.g., Fig~2 in][]{pipsee09}.
To take this into account, we modulate the value of $\mathcal{E}_{\phi}^{(W)}$,
given by Eq.~(\ref{eq:sh-cr}),
by  the quenching function $f_{4}^{(d)}\left(\Omega^{*}\right)$
which also appears in the expression for the $\vec{\Omega}\times\vec{J}$ effect
(penultimate term in Eq.~(\ref{eq:ephi}), proportional to $C_\delta^{(\Omega)}$);
additionally, the expression for the shear-current effect is normalized
so as to give Eq.~(\ref{difs1}) in the limit of slow rotation, for which one finds $f_{4}^{(d)}\left(\Omega^{*}\right)\approx(1/5)\,\Omega^{*}$.
Without such a quenching, i.e., directly applying Eq.~(\ref{difs1}), the shear-current effect would become unrealistically strong at the bottom of the convection zone.

In our numerical calculations we have used a dimensionless
form of 
the equations,
 substituting $r=xR_{\odot}$
and $t\rightarrow\eta_{0}t/R_{\odot}^{2}$,
where $\eta_0=1.8\cdot 10^9\,\mathrm{m}^2/\mathrm{s}$  is the maximum value of $\eta_T$ in the convection zone; that is, length is measured in units of the solar
radius and time is measured in units of the turbulent magnetic diffusion time, $T_D\approx8.5\, \mathrm{yr}$.
The integration
domain is radially bounded by $x=0.72$ and $x=0.96$.
The boundary conditions on the magnetic field are the usual approximate
perfect-conductor conditions, i.e., 
$A=0$, ${\displaystyle\frac{\partial xB}{\partial x}=0}$, at the bottom boundary
\citep{koh73},
and vacuum conditions, that is, $B=0$ and a continuous match
of the poloidal field component to an exterior potential field, at the top boundary.

The radial profiles of characteristic
quantities of the turbulence, such as the rms value $u_{c}$
and the
correlation length and time, $\ell_{c}$ and $\tau_{c}$, of the
convective background velocity field, as well as the density stratification,
were calculated
on the basis of a standard model of the solar interior \citep{sti02},
{assuming the ratio of the correlation length to the pressure
scale height (referred to as the mixing-length parameter) to be 1.6.}
The rotation profile as known from helioseismic measurements
\citep{schouetal98} is approximated by
\begin{equation}
\Omega=\Omega_{0}f\left(x,\theta\right),
\end{equation}
with
\begin{align}
f\left(x,\theta\right)=&\frac{1}{435}\left[435+51\left(x-x_{0}\right)+26\phi\left(x\right)\left(1-5\cos^{2}\theta\right) \right.\\
&\left.\quad \quad-3.5\left(1-14\cos^{2}\theta+21\cos^{4}\theta\right)\right] \,,
\end{align}
where
\begin{equation}
\phi\left(x\right)=0.5\left\{1+\mathrm{tanh}\left[50(x-x_{0})\right]\right\}
\end{equation}
and
$x_{0}=0.71$ is the position of tachocline, situated below the bottom
boundary of the integration domain.

{
A remark concerning our locating the lower boundary at $x=0.72$ seems in order.
Very often this boundary is placed at $x=0.65$ \citep[cf., e.g.,][]{jouetal08}.
Then, however, some modeling of the tachocline is needed, where the differential rotation
changes into rigid rotation in the radiative core. The physical parameters of this
transition region are rather uncertain at the moment. Here, we consider a convection-zone
dynamo model with distributed dynamo action in the bulk of the convection zone,
where all physical parameters needed can be derived from helioseismic measurements
and the standard model of the solar interior. There are other dynamo models where the
dynamo just operates in the tachocline \citep[a critical discussion of arguments for and against deep-seated and distributed dynamos is found in][]{bra05}.
}

The meridional flow, $\vec{U}$, is modeled in the form of two stationary circulation cells,
one in the northern and one in the southern hemisphere,
with poleward motion in the upper and equatorward motion in the lower part
of the convection zone \citep[for the theory of the meridional circulation see, e.g.,][]{rem05,rem06,mieetal08,brurem09}. The condition of mass conservation,
$\nabla\cdot(\rho\vec{U})=0$, is ensured by a stream-function representation
of $\rho\vec{U}$ \citep[cf., e.g.,][]{bonetal02}, so that
\begin{equation}\label{stream_function}
 \vec{U}=\frac{1}{\rho}\nabla\times\frac{\psi\vec{e}_\phi}{x\sin\theta}
=\frac{1}{\rho}\nabla\psi\times\frac{\vec{e}_\phi}{x\sin\theta} \,.
\end{equation}
The stream function, $\psi$, is written as
\begin{equation} \label{psi}
\psi(x,\theta) = u_0 \,\hat{\psi}(x,\theta)\,,
\end{equation}
with $u_0$ denoting the maximum amplitude of $\vec{U}$, which is treated as a free paramter.
Our choice for $\hat{\psi}(x,\theta)$ is
\begin{equation} \label{psi_hat}
\begin{split}
\hat{\psi}(x,\theta) =& c_0\sin^2\theta\cos\theta \, T(x) \,,
\\
&c_0=\left(\mathrm{max}\left|\frac{1}{\rho}\nabla\times\frac{\sin^2\theta\cos\theta \, T(x)\,\vec{e}_\phi}{x\sin\theta}\right|\,\right)^{-1} \,,
\end{split}
\end{equation}
where the function $T(x)$ is selected such that the top and bottom boundaries
are impenetrable and stress-free (free of tangential stresses), that is \citep[see, e.g.,][]{bat67},
\begin{equation}\label{stress-free}
 U_x=0, \quad
\frac{\partial U_x}{\partial \theta}+x\frac{\partial U_{\theta}}{\partial x} -U_{\theta}=0
\quad \mbox{at}\; x=x_{\mathrm{t}}, \, x=x_{\mathrm{b}}\,,
\end{equation}
where $x_{\mathrm{t}}$ and $x_{\mathrm{b}}$ are the outer and inner radius, respectively, of the considered spherical shell.
Using
 \begin{equation}
 U_{x} = \frac{u_{0}\,c_0\left(3\cos^{2}\theta-1\right)T}{\rho x^2} \,,\quad
 U_{\theta} = -\frac{u_{0}\,c_0\sin\theta\cos\theta}{\rho x}\frac{\mathrm{d} T}{\mathrm{d} x} \,,
\end{equation}
as follows from Eqs.~(\ref{stream_function}), (\ref{psi}), and (\ref{psi_hat}),
the boundary conditions given by Eq.~(\ref{stress-free}) take the form
\begin{equation} \label{stress-free-2}
 T=0 \,, \quad x\frac{\mathrm{d}}{\mathrm{d} x}\frac{1}{\rho x}\frac{\mathrm{d} T}{\mathrm{d} x}
-\frac{1}{\rho x}\frac{\mathrm{d} T}{\mathrm{d} x}=0
\quad \mbox{at}\; x=x_{\mathrm{t}}, \, x=x_{\mathrm{b}}
\end{equation}
and are satisfied with
\begin{equation}
T\left(x\right)=  P_{2}\left(\xi\right)+c_{1}\left[P_{3}\left(\xi\right)-\xi\right]+c_{2}\left[P_{4}\left(\xi\right)-1\right]-1 \,,
\end{equation}
where
\begin{equation}
\xi = \frac{2x-\left(x_{\mathrm{t}}+x_{\mathrm{b}}\right)}{x_{\mathrm{t}}-x_{\mathrm{b}}}
\end{equation}
is the radial coordinate transformed from the interval $[x_{\mathrm{b}},x_{\mathrm{t}}]$
to the interval $[-1,1]$,
$P_{n}$  denotes the Legendre polynomial of degree $n$,
 and
$c_{1}\approx-0.207$ and $c_{2}\approx-0.097$
are numerically determined constants; for the normalization constant $c_0$ (cf. Eq.~(\ref{psi_hat}))
one then finds
$c_0\approx 0.027$.

\begin{figure*}
\begin{centering}
\includegraphics[width=0.25\textwidth,height=0.25\textwidth]{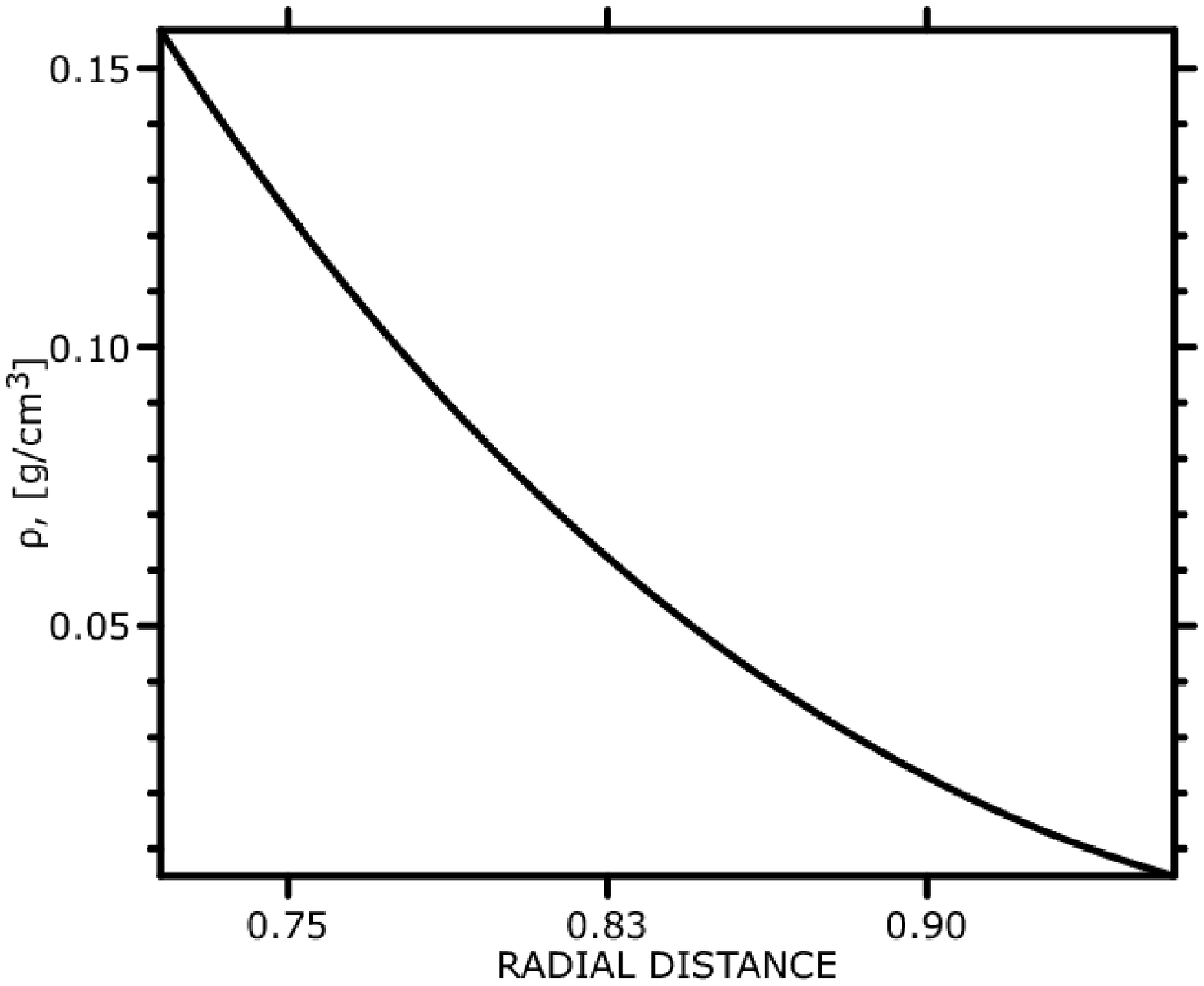}
\hspace{1em}
\includegraphics[width=0.5\textwidth,height=0.25\textwidth]{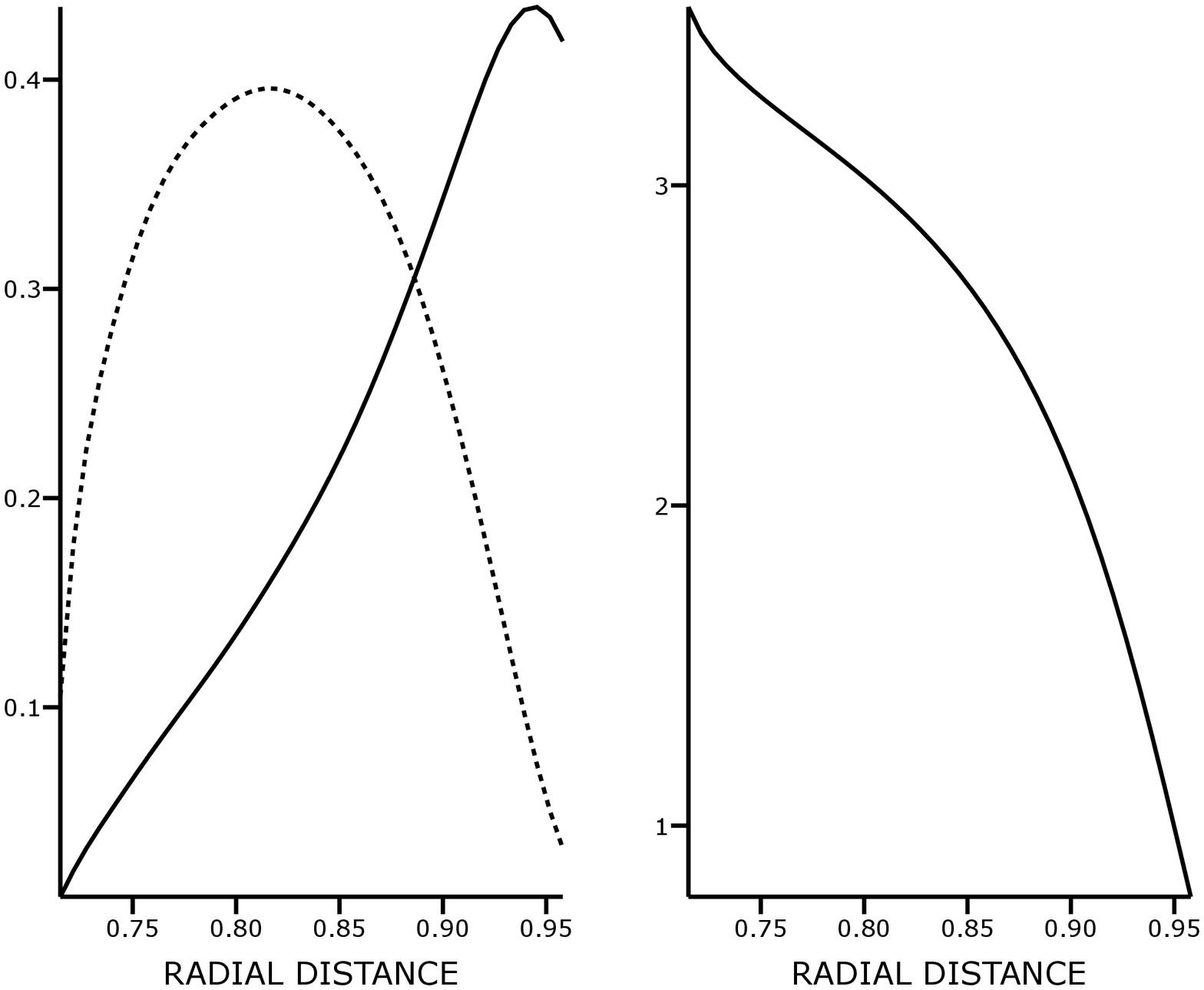}
\caption{\label{fig:profiles} Radial profiles of model quantities.
{{\em Left:} Mass density.}
{\em Middle:} Isotropic
($\eta_T f_{2}^{(d)}\left(\Omega^{*}\right)$, solid line) and anisotropic
($2\eta_T f_{1}^{(a)}\left(\Omega^{*}\right)$, dashed line) 
turbulent magnetic diffusivities in units of $\eta_0$. {\em Right:}
Effective strength of the $\vec{\Omega}\times\vec{J}$ effect,
$f_{4}^{(d)}\left(\Omega^{*}\right)/\left(f_{2}^{(d)}
\left(\Omega^{*}\right)+2f_{1}^{(a)}\left(\Omega^{*}\right)\right)$.}
\end{centering}
\end{figure*}

\begin{figure}
\includegraphics[width=0.8\linewidth]{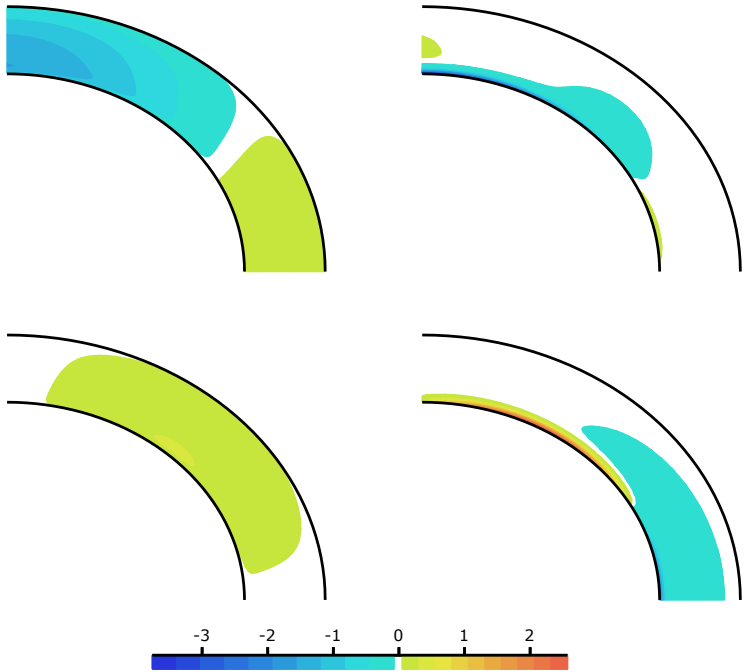}
\caption{\label{fig:shear-current}
Effective strengths of the contributions to the shear-current effect as
given by Eq.~(\ref{eq:sh-cr}):
$(11/6)\left(\hat{\Omega}-1\right)N$ {\em (top left)},
$(1/6)\left(\cos\theta\,\partial\hat{\Omega}/\partial x-
(\sin\theta/x)\,\partial\hat{\Omega}/\partial\theta\right)N$ {\em (top right)},
$(1/3)\sin\theta\left(\partial\hat{\Omega}/\partial\theta\right)  N$ {\em (bottom left)},
and $-(1/3)\sin\theta\left(\partial\hat{\Omega}/\partial r\right)  N$ {\em (bottom right)},
with
$N=f_{4}^{(d)}\left(\Omega^{*}\right)/\left(f_{2}^{(d)}\left(\Omega^{*}\right)+2f_{1}^{(a)}\left(\Omega^{*}\right)\right)$.
}
\end{figure}

\begin{figure}
\includegraphics[width=0.8\linewidth]{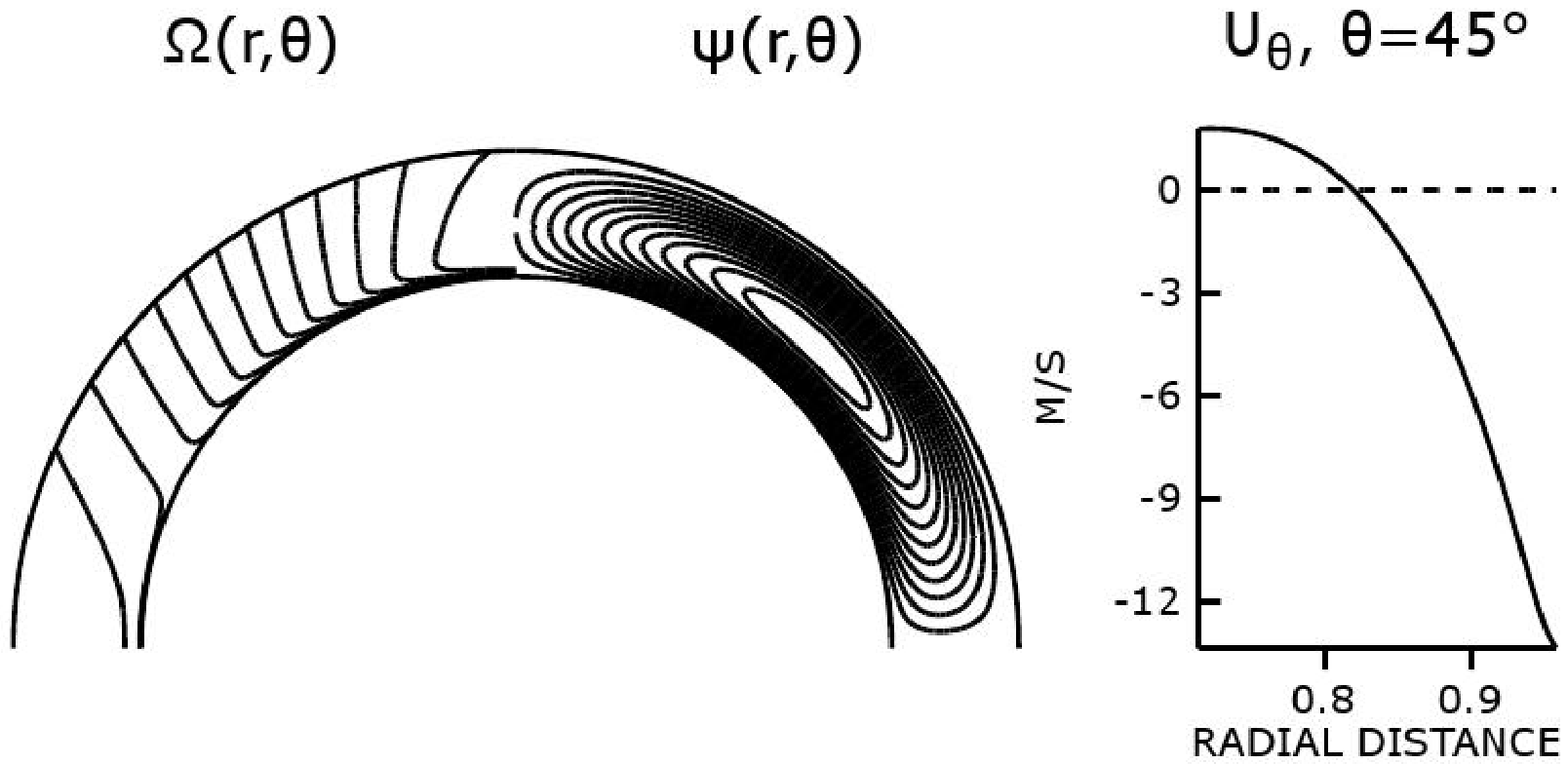}
\caption{
\label{fig:flows}The large-scale flows. {\it Left panel:} Contours of the rotation rate $\Omega$ {\it (left)} and
streamlines of the meridional flow, $\vec{U}$, i.e., contours of its
stream function, $\psi$  {\it (right)}.
{\it Right panel:} Radial profile of the component $U_\theta$  of the
meridional flow at a latitude of $45^\circ$ (in m/s).
}
\end{figure}

Fig.~\ref{fig:profiles}
 shows radial profiles of
{the mass density,} the isotropic
and anisotropic magnetic diffusivities, and  the effective strength of the
$\vec{\Omega}\times\vec{J}$ effect,
and in Fig.~\ref{fig:shear-current}
 the effective strengths of the
contributions to the shear-current effect as given by Eq.~(\ref{eq:sh-cr})
are displayed.
The properties of the two large-scale flows used in the model, namely, contours of the differential rotation rate and the
geometry and strength of the meridional circulation, are depicted in Fig.~\ref{fig:flows}.
Other model quantities, such as the radial profile of the Coriolis number,
$\Omega^{\ast}$, and the pumping velocity of the toroidal part of the LSMF, are
as in \citet[][see Fig.~2 there]{pipsee09}.

\subsection*{Numerical procedure}

In our numerics we use a Galerkin method, expanding the magnetic field in terms
of a basis that satisfies the boundary conditions implicitly.
The system of  Eqs.~(\ref{eq:2}) and (\ref{eq:1}) admits exponentially
growing or decaying solutions, which we
represent in the form
 \begin{align}
A\left(x,\theta,t\right) & =  \mathrm{e}^{{\displaystyle
      \lambda t}}\sum_{n}\sum_{m}A_{nm}
\sin\theta \, S_{nm}^{(A)}\left(\xi\right)P_{m}^{1}\left(\cos\theta\right),\label{eq:A-dec} \\
B\left(x,\theta,t\right) & =  \mathrm{e}^{\displaystyle{\lambda t}}\sum_{n}\sum_{m}B_{nm}S_{n}^{(B)}\left(\xi\right)P_{m}^{1}\left(\cos\theta\right),\label{eq:B-dec}
\end{align}
where $S_{nm}^{(A)}$ and $S_{n}^{(B)}$ are linear combinations of Legendre polynomials, and
 $P_{m}^{1}$ is the associated Legendre function of degree $m$ and order $1$.
These expansions ensure the regularity of the solutions at the poles
$\theta=0$ and $\theta=\pi$,
where $B_{\theta}$ and $B_{\phi}$, i.e., $\displaystyle \frac{A}{\sin\theta}$ and $B$,
have to vanish.
The radial boundary conditions are satisfied by the choice \citep[see ``basis recombination'' in ][]{boy01,livjac05}
\begin{align}
S_{nm}^{(A)}\left(\xi\right)& =  P_{n-1}\left(\xi\right)+a^{(1)}_{nm}P_{n}\left(\xi\right)+a^{(2)}_{nm}P_{n+1}\left(\xi\right),\label{eq:A-dec1}\\
S_{n}^{(B)}\left(\xi\right) & =  P_{n-1}\left(\xi\right)+b^{(1)}_nP_{n}\left(\xi\right)+b^{(2)}_nP_{n+1}\left(\xi\right),\label{eq:B-dec1}
\end{align}
where
\begin{align}
a^{(1)}_{nm}  =  \frac{2n+1}{(n+1)^{2}+2m/\gamma}\,, &\quad  a^{(2)}_{nm}=-\frac{n^{2}+2m/\gamma}{(n+1)^{2}+2m/\gamma}\,, \label{a_nm}\\
b^{(1)}_n  =  -\frac{x_{\mathrm{b}}(2n+1)}{x_{\mathrm{b}}(n+1)^2-2/\gamma}\,, &\quad  b^{(2)}_n=-\frac{x_{\mathrm{b}}n^{2}-2/\gamma}{x_{\mathrm{b}}(n+1)^2-2/\gamma}\,,\label{b_n}
\end{align}
with $\gamma={\displaystyle \frac{2}{x_{\mathrm{t}}-x_{\mathrm{b}}}}$ denoting the derivative of $\xi$ with
respect to $x$.

Integrations over radius and latitude, necessary for calculating the expansion coefficients
$a_{nm}$ and $b_{nm}$, were done by means of the Gauss-Legendre
procedure, and the eigenvalue problem for determining the exponent $\lambda$ and the associated
eigenmodes was
solved by means of Lapack routines.
{The spectral resolution was 15 modes in the radial basis and 22 modes in the latitudinal basis for the 
calculations of growth rates (including stability boundaries) and dynamo periods, and $14\times 30$ modes for simulations of time evolutions and butterfly diagrams (see Sect.~\ref{sec_results} below);
 by the assumption of either dipole-type or quadrupole-type solutions the latitudinal resolution could be doubled in a part of the calculations. The results
were qualitatively confirmed by a number of runs with still higher resolution. 
Benchmark calculations for the code used are presented in Appendix \ref{appendix_benchmarks}.}

\section{Results}
\label{sec_results}

\subsection{$\delta^{(\Omega)}\Omega$ dynamo with meridional flow}
\label{sec_delta-Omega-dynamo}

The meridional flow becomes essential for the dynamo if the effective
magnetic Reynolds number, based on the meridional flow velocity and the turbulent
magnetic diffusivity, is high enough. Meridional flow velocities higher than about
$10\dots 20$ m/s can scarcely be brought into agreement with the solar observations.
Thus, the turbulent
magnetic diffusivity should be low. In our formulation, all  turbulence
effects are consistently scaled by the parameter $C_{\eta}$.
{Decreasing  $C_{\eta}$ leads to increasing  the influence of the flow on the magnetic field and
acts, thus, like increasing the amplitude of the flow.}
 An advection-dominated
regime with a solar-like magnitude of the meridional flow is obtained if
$C_{\eta}\lessapprox 1/20$. Below, we fix $C_{\eta}$ to the value {$1/40$}.

\begin{figure}
\begin{centering}
\includegraphics[angle=-90,width=0.5\textwidth]{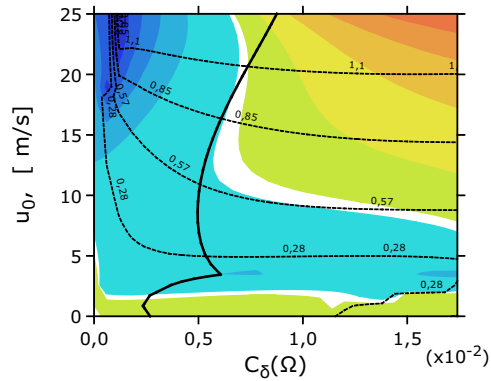}
\par\end{centering}
\centering{}\caption{\label{fig3-1}Difference between the growth
rates of the first (most unstable) dipolar mode and the first
quadrupolar mode in the plane spanned by $C_{\delta}^{(\Omega)}$ and $u_0$ for $C_{\delta}^{(W)}=0$.
 Red colors correspond to a
dominance of dipolar modes and blue colors to a dominance of quadrupolar modes.
Also shown are isolines of the frequency of the first dipolar mode (in areas where it oscillates),
with numbers giving the oscillation frequency ($2\pi/T$, $T$ being the time period of the
oscillation) in units of the inverse magnetic
diffusion time, $\eta_{0}/R_{\odot}^2$.
The solid bold line indicates the stability boundary for the first dipolar mode,
with the unstable (dynamo) region lying to the right of this line.
}
\end{figure}

Figure \ref{fig3-1}
illustrates the bifurcation scenario for a $\delta^{(\Omega)}\Omega$ dynamo
 with meridional flow. Here the sources of the dynamo are the
$\vec{\Omega}\times\vec{J}$ effect (the shear-current effect being neglected) and differential rotation. Also included are turbulent diffusion (isotropic and anisotropic) and turbulent pumping. As the oberservations of solar activity suggest,
the large-scale solar magnetic field is characterized by an antisymmetric
parity with respect to the equatorial plane.
In the figure, the plane spanned
by the parameters $C_\delta^{(\Omega)}$ (measuring
the strength of the $\vec{\Omega}\times\vec{J}$ effect)
and $u_0$ (the maximum amplitude of the meridional flow) is displayed,
with red/blue colors
indicating a dominance of dipolar/quadrupolar modes, which are antisymmetric/symmetric with respect to the equatorial plane, over quadrupolar/dipolar modes.
For a sufficiently high velocity of the meridional flow 
the dipolar parity dominates,
as needed for the Sun.
{The critical dynamo number $C_\delta^{(\Omega)}$ where the first
dipolar mode becomes unstable does not depend very much on the flow strength.
This may result from the fact that the meridional flow mainly acts as a conveyor belt
for the magnetic field, rather than as a generation mechanism.
}

\begin{figure}
\begin{centering}
\includegraphics[angle=-90,width=0.5\textwidth]{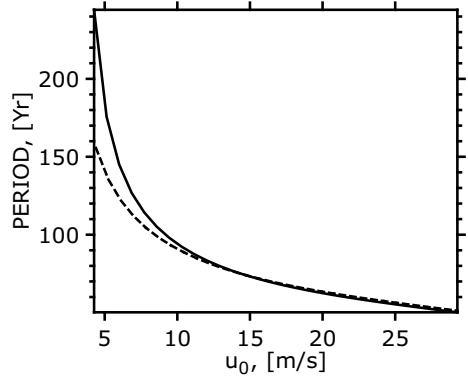}
\par\end{centering}
\centering{}\caption{\label{period}{Dependence of the dynamo period on $u_0$
along the stability boundary of the most unstable dipolar mode
for the $\delta^{(\Omega)}\Omega$ (solid line) and $\delta^{(W)}\Omega$
(dashed line)
dynamos.}
}
\end{figure}

{In Fig.~\ref{period} (solid line)}
{the dependence of the dynamo period on the amplitude of the meridional flow
along the stability boundary is shown. As expected, the period is a decreasing
function of the flow velocity \citep[cf.][]{bonetal02,bonelsbel06}. The dependence approximately follows a power law with a scaling exponent of $-0.7$.}

\begin{figure*}
\begin{centering}
\includegraphics[angle=-90,width=0.8\textwidth]{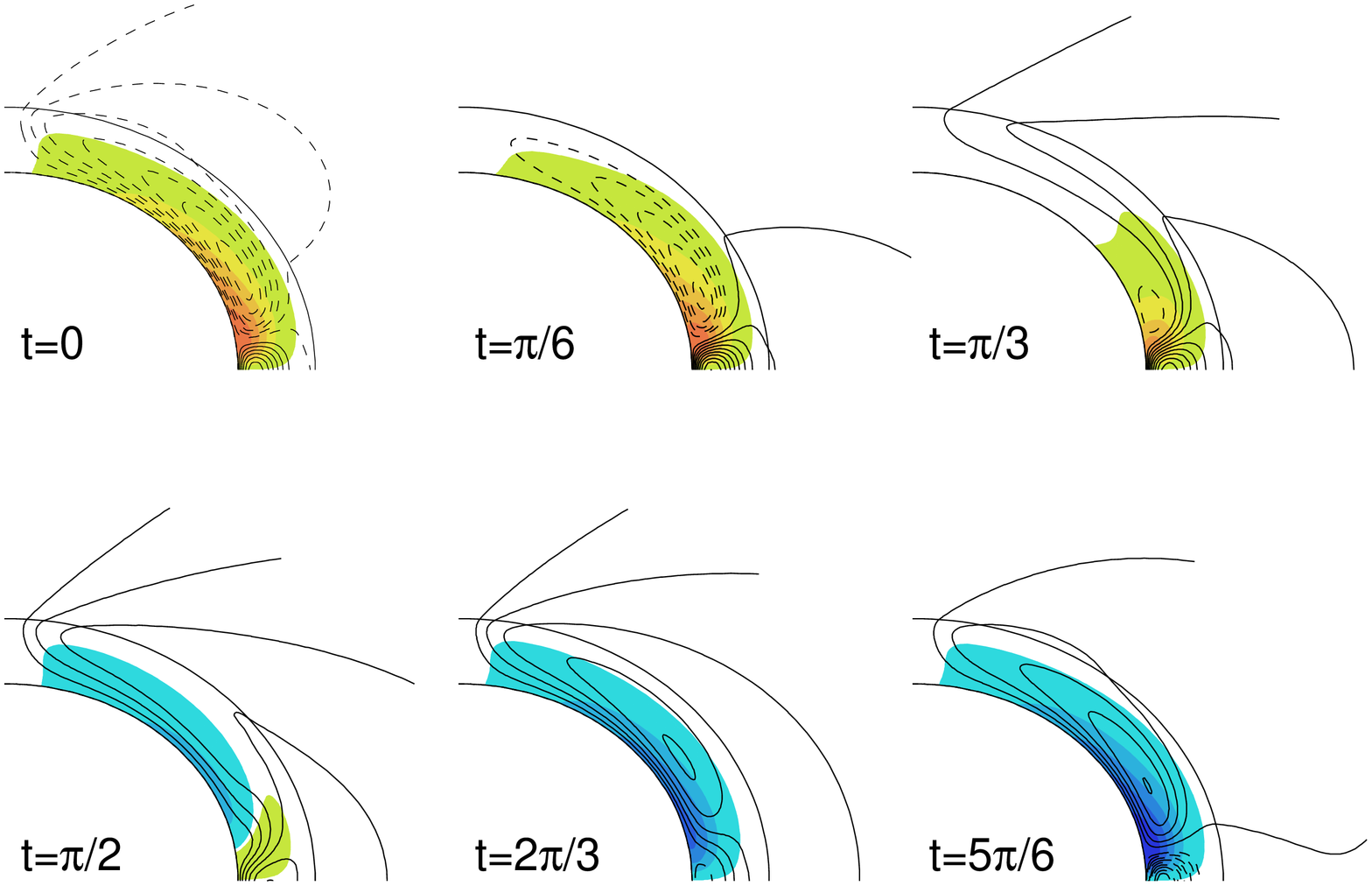}
\par\end{centering}
\begin{centering}
\includegraphics[width=0.8\textwidth]{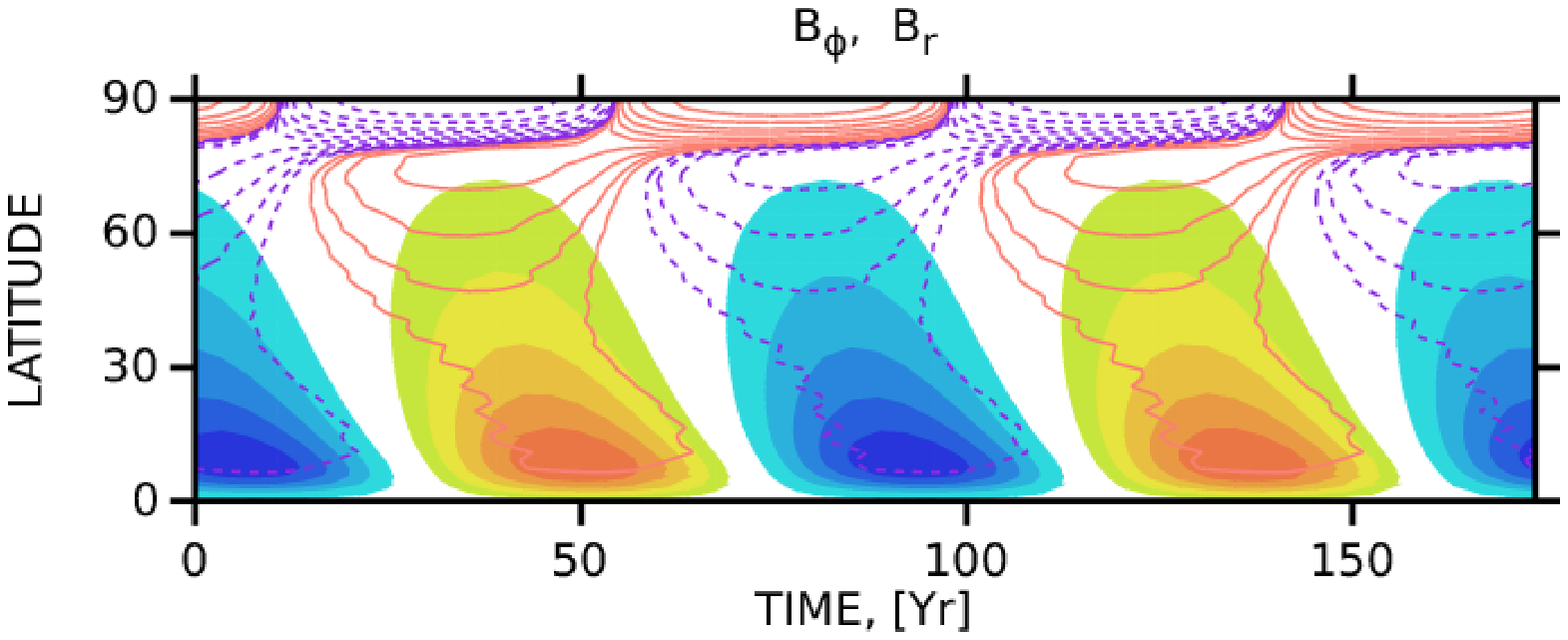}
\par\end{centering}
\centering{}\caption{\label{fig1}
$\delta^{(\Omega)}\Omega$ dynamo for $C_{\delta}^{(\Omega)}=0.006$ and $u_0=15\, \mathrm{m/s}$.
{\it Top and middle:} Snapshots of the strength of the toroidal LSMF (color-coded, red/blue colors correspond to positive/negative field values) and field lines of the
poloidal LSMF (solid/dashed lines indicate clockwise/counter-clockwise field direction) over half a cycle in the meridional plane.
{\it Bottom:} Associated butterfly diagram, showing the (color-coded) strength of the toroidal LSMF (integrated over depth in the convection zone) in the time-latitude plane.
Also shown are isocontours of the radial field component at the top boundary
(solid red/dashed blue lines indicate positive/negative values).}
\end{figure*}

Figure \ref{fig1} (top and middle)
shows, in the form of the strength of the toroidal field
and field lines of the poloidal field in the meridional plane, the evolution of the LSMF in an $\delta^{(\Omega)}\Omega$ dynamo
model on the basis of the first dipolar mode for 
{$C_{\delta}^{(\Omega)}=0.006$ and $u_0=15\,\mathrm{m/s}$}.
In the bottom panel of the figure, an associated simulated butterfly diagram,
i.e., the strength of the toroidal LSMF (integrated over depth in the convection zone) in the time-latitude plane is shown, together with isocontours
of the radial field component at the top boundary of the considered spherical shell.
The latitudinal drift of the toroidal field towards the equator, as indicated by the observations of sunspots, is qualitatively correctly reproduced;
the poloidal field is coupled to the toroidal field and shows a similar drift
towards the equator. 
The phase relation between the toroidal and poloidal parts
of the magnetic field, in particular
the polar reversal of $B_{r}$ shortly after the maximum
of $B_{\phi}$ at low latitudes, with $B_rB_\phi<0$ before and $B_rB_\phi>0$
after the polar field reversal, is also in good agreement with the solar observations.

In the example shown in Fig.~\ref{fig1}, the obtained cycle period is about
{ $80\,\mathrm{yr}$}, which is nearly
{four} times the period of the solar activity cycle. Tuning the paramters cannot significantly
reduce the period.
This may appear not fully satisfactory,
but the period is at least in the right order of magnitude.
Furthermore, for the example in Fig.~\ref{fig1}, we find 
{$B_\phi/B_r\approx 15$, which is by a factor
of $10\dots 100$} smaller than the ratio between the large-scale toroidal and poloidal fields
as supposed for the solar convection zone. In general,
increasing the speed of the meridional flow in the model reduces the obtained ratio between the toroidal and poloidal fields, in apparent conflict with the need to reduce the cycle period.

\subsection{$\delta^{(W)}\Omega$ dynamo with meridional flow}
\label{sec_delta-W-dynamo}

\begin{figure}
\begin{centering}
\includegraphics[angle=-90,width=0.5\textwidth]{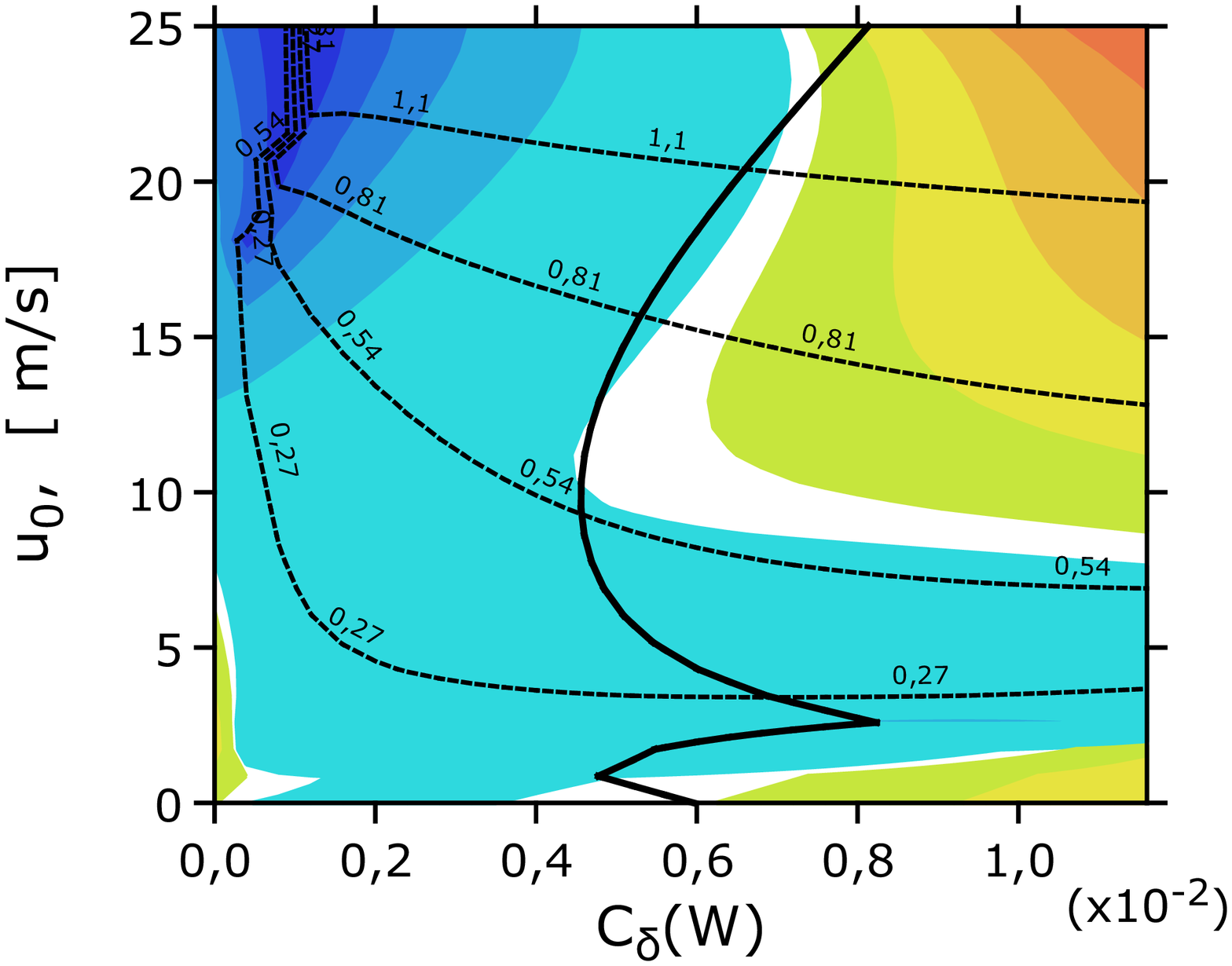}
\par\end{centering}
\centering{}\caption{\label{fig4-1-1}
As Fig.~\ref{fig3-1}, but with $C_\delta^{(\Omega)}$ replaced by $C_\delta^{(W)}$,
and $C_\delta^{(\Omega)}=0$.}
\end{figure}

\begin{figure*}
\begin{centering}
\includegraphics[angle=-90,width=0.8\textwidth]{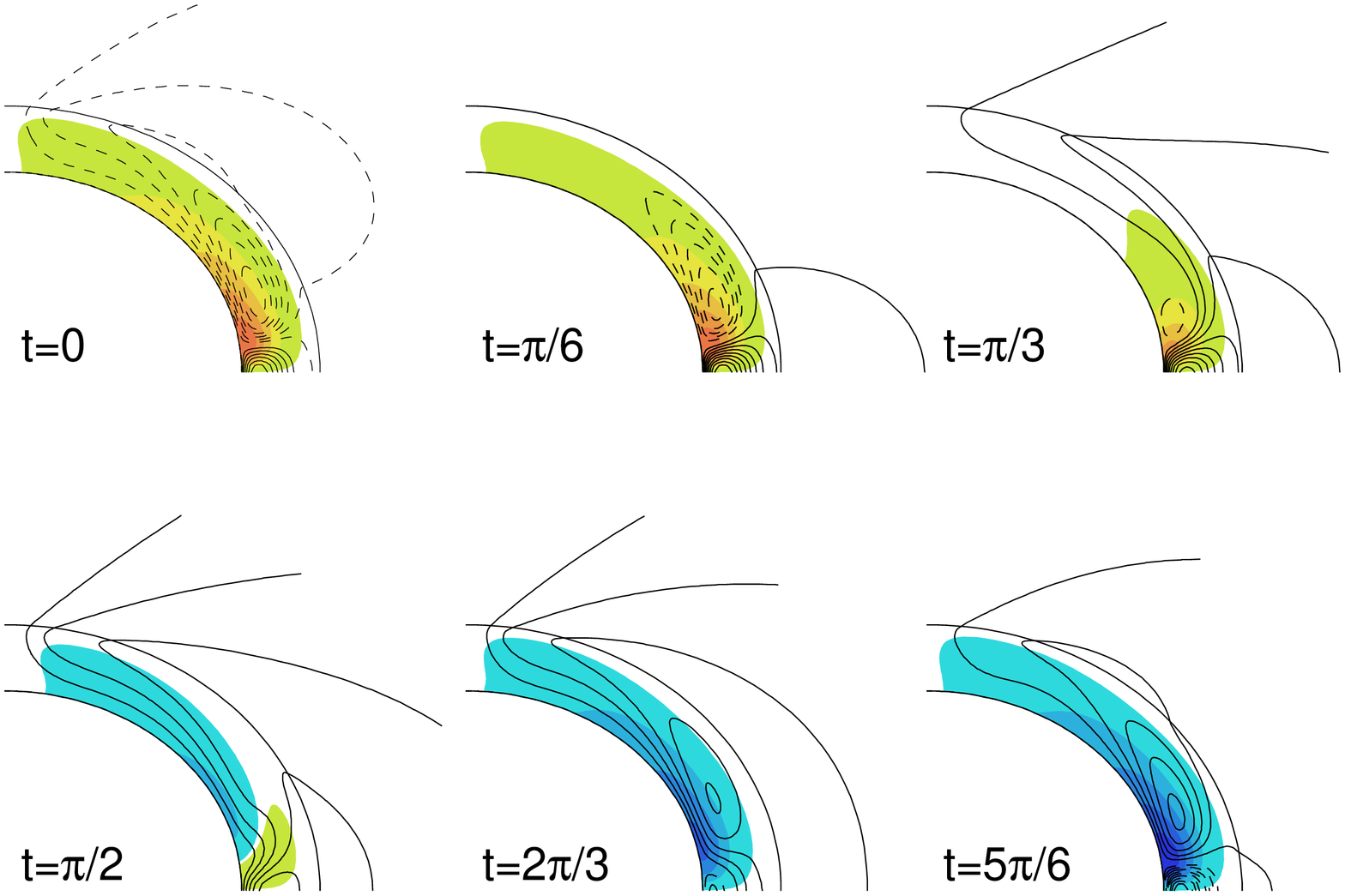}
\par\end{centering}
\begin{centering}
\includegraphics[width=0.8\textwidth]{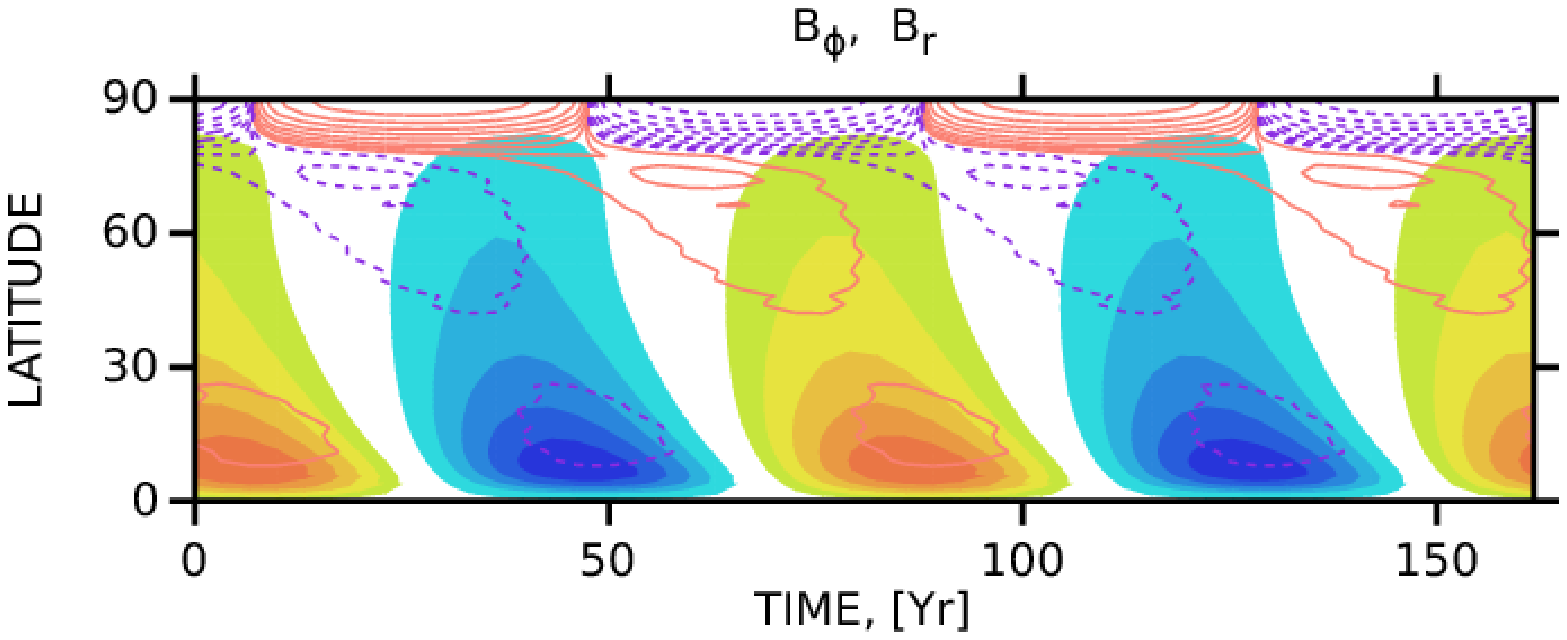}
\par\end{centering}
\centering{}\caption{\label{fig2}
As Fig.~\ref{fig1}, but for a
$\delta^{(W)}\Omega$ dynamo with $C_{\delta}^{(W)}=0.007$ and $u_0=15\, \mathrm{m/s}$.}
\end{figure*}

Figures \ref{fig4-1-1} and \ref{fig2}
show results for a dynamo on the basis of the shear-current effect, differential
rotation, and a meridional
flow, with the $\alpha$ effect and the $\vec{\Omega}\times\vec{J}$ effect being neglected
(but again with turbulent diffusion and turbulent pumping being included).
The results strongly resemble those for the $\delta^{(\Omega)}\Omega$ dynamo with meridional flow as described in Sect.~\ref{sec_delta-Omega-dynamo},
obviously due to the similarity between the $\vec{\Omega}\times\vec{J}$ effect
and the shear-current effect.
The comments given in Sect.~\ref{sec_delta-Omega-dynamo} thus apply here as well.
{The dependence of the dynamo period on the amplitude of the meridional flow
along the stability boundary (Fig.~\ref{period}, dashed line) approximately follows a power law with a scaling exponent of $-0.6$.}

\section{Conclusions}
\label{sec_conclusions}

We have studied kinematic axisymmetric mean-field dynamo
models  in the geometry of a spherical shell, as appropriate for the Sun and
solar-type stars, where the $\vec{\Omega}\times\vec{J}$
and shear-current effects were included as turbulent sources of
the large-scale magnetic field while the $\alpha$ effect was omitted.
Besides the turbulent dynamo mechanisms and differential rotation, a meridional
circulation, in the form of two stationary circulation cells,
one in the northern and one in the southern hemisphere, also
was incorporated into the models. We have concentrated on the dynamo
onset.

Our results show that the $\vec{\Omega}\times\vec{J}$ and shear-current effects can,
at least in principle,
take over the role that the $\alpha$ effect usually plays in mean-field
dynamo models. However, only if the meridional flow is sufficiently fast
are the characteristic properties of solar-type dynamos qualitatively
correctly reproduced. In particular, the amplitude of the meridional flow
needs to exceed a threshold value in order that the most unstable magnetic mode has dipolar parity and oscillates. This mode then also shows a latitudinal drift
towards the equator within each half cycle and a phase relation between the
poloidal and toroidal parts of the field in accordance with the observations
of solar activity. The threshold value for the amplitude of the meridional flow,
$u_0\gtrsim 10\,\mathrm{m/s}$ (reached at the surface, the flow speed at the bottom is on the order of
 $1\,\mathrm{m/s}$), is consistent with solar
observations and agrees with the value of $10\,\mathrm{m/s}$ often adopted in
studies of flux-transport dynamos with the $\alpha$ effect \citep[cf., e.g.,][]{bonetal02,jouetal08}.

{
In models of advection-dominated dynamos, the specifics of the turbulent dynamo mechanism that
generates the mean poloidal field are less important than they are in models without meridional
flow (given the rotation law and the generation of the mean toroidal field from the
mean poloidal field by velocity shear). Once the field is generated, it is advected equatorwards by the flow. However, the distribution of the turbulent dynamo sources,
or their more or less strong localization, decisively influences the parity properties
of the LSMF.}
Studies of flux-transport dynamos with an $\alpha$ effect as the turbulent source of
the LSMF indicate that the $\alpha$ effect must be strongly localized at the bottom of 
the convection zone to ensure the correct (dipolar) parity of the LSMF
\citep{dikgil01,bonetal02}. In our models, the turbulent dynamo effects are
distributed over the bulk of the convection zone, though they are strongest near the bottom of the included domain. 
{We note that our  turbulent dynamo sources
are not introduced arbitrarily but are calculated using a standard model of the
solar interior together with rotation rates obtained from helioseismic measurements.
}

As other advection-dominated dynamo models, the models presented here work only if the
effective magnetic diffusivity is strongly reduced compared to the mixing-length estimates. At a radial distance of, say, $0.85\,R_{\odot}$, the turbulent magnetic diffusivity in our models is about $0.2\,\eta_0$ 
($\eta_0=1.8\cdot 10^9\,\mathrm{m}^2/\mathrm{s}$
is the maximum value of the turbulent magnetic diffusivity in the convection zone
according to the mixing-length estimate). Together with our value of {0.025}
for the parameter $C_\eta$ (which regulates the turbulence level), this gives
an effective magnetic diffusivity of about
{$10^7\,\mathrm{m}^2/\mathrm{s}$},
in agreement with the upper limit
of $3\cdot 10^7\,\mathrm{m}^2/\mathrm{s}$ for the turbulent magnetic diffusivity
in the bulk of the convection zone given by \citet{dikgil06,dikgil07} for
flux-transport dynamos with the $\alpha$ effect.

The cycle periods that we obtain are at least {three times} as long as the observed period of the solar
activity cycle. Also, the ratio between the toroidal and poloidal parts of the large-scale
magnetic field is significantly smaller than supposed for the solar convection zone
(apparently a common problem of all flux-transport models). Here one should keep in mind that requiring
a perfect fit to the solar details, as far as these are known, would overstress the models.
For instance, a solution that bifurcates at the dynamo onset will change quantitatively
if it is traced away from the bifurcation point.
{Thus, ultimately, self-consistent nonlinear models will be needed.}

\begin{acknowledgements}
The work of V.~V.~Pipin was supported by the Russian Foundation for Basis Research
(RFBR) through grants 07-02-00246, 2258.2008.2, and 09-02-91338.
\end{acknowledgements}


\newpage
\begin{appendix}

\section{Benchmarks for the code used}
\label{appendix_benchmarks}

Here we present some benchmark tests for the computer code that we used
{and show how the crititical dynamo numbers and dynamo
periods for the models of Sects.~\ref{sec_delta-Omega-dynamo}
and \ref{sec_delta-W-dynamo} converge for an increasing number of modes taken into
account.}

\subsection{Free decay modes}

First we test the accuracy of the implementation of the exterior boundary conditions
and the speed of convergence. If we neglect
all the dynamo effects in Eqs. (\ref{eq:2}) and (\ref{eq:1}), only simple isotropic
diffusion remains and the equations take the form
\begin{align}
\frac{\partial A}{\partial t} & =  \frac{\partial^{2}A}{\partial x^{2}}+\frac{\sin\theta}{x^{2}}\frac{\partial}{\partial\theta}\frac{1}{\sin\theta}\frac{\partial A}{\partial\theta} \,,\label{A:eq:2}\\
\frac{\partial B}{\partial t} & =  \frac{1}{x}\frac{\partial^{2}\left(xB\right)}{\partial x^{2}}+\frac{1}{x^{2}}\frac{\partial}{\partial\theta}\frac{1}{\sin\theta}\frac{\partial\left(\sin\theta B\right)}{\partial\theta} \,,\label{A:eq:1}
\end{align}
where for simplicity the magnetic diffusivity has been assumed
to be homogeneous and has been set equal to unity. The equations for the poloidal
and toroidal parts of the field are decoupled here. The eigenmodes to
Eqs. (\ref{A:eq:2}) and (\ref{A:eq:1}) are the free decay modes,
exponentially decaying $\propto \exp{\lambda_i t}$, where the $\lambda_i$ are the
eigenvalues of the Laplacian operator for the considered domain under the
imposed boundary conditions; these eigenvalues are all real and negative.
For the test, we consider the case of a full sphere (rather than a spherical shell)
surrounded by vacuum, for which
the free decay modes can be determined analytically and are well documented in the literature
\citep[see, e.g.,][]{mof78,bacparcon96}.
For that purpose,  the potential functions $A$ and $B$
are written as
\begin{align}
A\left(x,\theta,t\right) & =  \mathrm{e}^{{\displaystyle
      \lambda t}}\sum_{n}\sum_{m}A_{nm}
\sin\theta \, S_{nm}^{(A)}\left(x\right)P_{m}^{1}\left(\cos\theta\right),\label{eq:A-sphere} \\
B\left(x,\theta,t\right) & =  \mathrm{e}^{\displaystyle{\lambda t}}\sum_{n}\sum_{m}B_{nm}S_{n}^{(B)}\left(x\right)P_{m}^{1}\left(\cos\theta\right),\label{eq:B-sphere}
\end{align}
with
\begin{align}
S_{nm}^{(A)}\left(x\right) & =  x\left(P_{2n+1}\left(x\right)-P_{1}\left(x\right)\frac{\left(2n+1\right)\left(2n+2\right)+2\left(m+1\right)}{2m+4}\right),\label{N:eq:A-dec1} \\
S_{n}^{(B)}\left(x\right) & =  x\left(P_{2n+1}\left(x\right)-P_{1}\left(x\right)\right),\label{N:eq:B-dec1}
\end{align}
where the radial variable, $x$, varies in the interval
$\left[0,1\right]$; the transformation to the variable $\xi$ (cf. Eqs. (\ref{eq:A-dec})
and (\ref{eq:B-dec}) in Sect. \ref{sec_model}) is not used here.
By the choice of the basis functions given by Eqs. (\ref{N:eq:A-dec1}) and
(\ref{N:eq:B-dec1}) the exterior vacuum conditions are satisfied and the regularity of the fields at the origin is ensured. This set of basis functions
differs from that used in our calculations for the spherical shell, but the general
structure of the code and the solution algorithms are not changed.

The dependence of the solutions to Eqs. (\ref{A:eq:2}) and (\ref{A:eq:1}) on radius
is given analytically in terms of the spherical Bessel functions $j_n(x)\propto \left(1/\sqrt{x}\right)J_{n+1/2}(x)$ (where $J_{n+1/2}$ is the ordinary Bessel function
of half-integer order $n+1/2$), and the decay rates, $-\lambda_n$, are given by the squares of the zeros of the functions 
$j_{n-1}$ for the poloidal and $j_n$ for the toroidal modes.
The slowest decaying (largest scale) poloidal mode decays with the rate
$-\lambda_1^{(A)}=\pi^2$, the slowest decaying toroidal mode with the rate
$-\lambda_1^{(B)}\approx 4.4934^2$;
the corresponding eigenfunctions are $S_1^{(A)}(x,\theta)\propto j_1\left(\lambda_1^{(A)}x\right)\sin^2\theta$
and $S_1^{(B)}(x,\theta)\propto j_1\left(\lambda_1^{(B)}x\right)\sin\theta$.
 These two decay modes are used for the test.
The $\theta$ dependences of their potential functions
are given by the first terms (with $m=1$) in the latitudinal expansions
on the right-hand sides of Eqs.\ (\ref{eq:A-sphere}) and  (\ref{eq:B-sphere}).
(The potential functions $A$ and $B$ differ from the potentials $S$ and $T$ in the
poloidal-toroidal decomposition
$\vec{B} =\nabla\times(\vec{r}\times\nabla S) + \vec{r}\times\nabla T$
as normally used in non-axisymmetric cases,
see, e.g., \citet{mof78} and \citet{bacparcon96}. For our axisymmetric case,
one has
$\partial S/\partial\theta = A/\sin\theta$ and $\partial T/\partial\theta = B$.
The angular dependence of both $S$ for the slowest decaying poloidal
mode and $T$ for the slowest deacaying toroidal mode is given by
the spherical surface harmonic $Y_1^0(\cos\theta)\propto \cos\theta$,
in agreement with the $\theta$ dependences of $A$ and $B$ as given above.)

Table \ref{table_decay-modes}
\begin{table}
\caption{\label{table_decay-modes} Convergence of the eigenvalues and eigenvectors
 of the slowest decaying poloidal
and toroidal modes. $^{\mathrm{a}}$}
\begin{centering}
\begin{tabular}{ccccc}
\hline 
$N$ &$E(\lambda)\,[A]$ & $E(\vec{B})\,[A]$& $E(\lambda)\,[B]$ &$E(\vec{B})\,[B]$\tabularnewline
\hline
3 &1.08e-7 & 3.98e-9& 3.83e-5 & 6.849e-4\tabularnewline
4 &5.651e-11 & 1.255e-12& 9.984e-8 & 4.365e-9\tabularnewline
5&9.68e-14 & 2.142e-16 & 1.207e-10 & 2.497e-12\tabularnewline
6 &7.72e-14 & 2.136e-20& 8.707e-14 & 9.068e-16\tabularnewline
7 &1.38e-14 & 1.325e-24& 1.77e-14 & 4.80e-19\tabularnewline
8 &1.90e-14 & 4.427e-29& 5.32e-15 & 6.263e-23\tabularnewline
\hline
\end{tabular}
\par\end{centering}
\begin{list}{}{}
\item[$^{\mathrm{a}}$]
$N$ is the number of modes in the radial basis,
and $E(\lambda)\,[A]$ and $E(\vec{B})\,[A]$ are the errors of eigenvalue and eigenvector for the poloidal mode and $E(\lambda)\,[B]$ and $E(\vec{B})\,[B]$
the corresponding errors for the toroidal mode.
\end{list}
\end{table}
shows the convergence of the eigenvalues and of
the corresponding eigenvectors for our numerical scheme.
Similar to \cite{livjac05}, the eigenvectors
are scaled so that $S_1^{(A)}\left(x=1,\theta=\pi/2\right)=1$ and $S_1^{(B)}\left(x=0.5,\theta=\pi/2\right)=1$,
and the errors are measured as $E\left(\lambda\right)=\left|\lambda_{\mathrm{true}}-\lambda_{\mathrm{num}}\right|$
and $E\left(\vec{B}\right)=\int_V\left|\vec{B}_{\mathrm{true}}-\vec{B}_{\mathrm{num}}\right|^{2}\mathrm{d}V$.
The number of modes in the radial basis, $N$, is varied, while
in the latitudinal basis just the first mode is taken into account.
The convergence is seen to be exponential in both the poloidal and toroidal cases.

\subsection{Test case B of \citet{jouetal08}}

The next test case is taken from \citet{jouetal08}, who presented a comparitative
benchmark study of different numerical codes for axisymmetric mean-field solar
dynamo models in spherical geometry. Here we consider their test case B,
which is a pure $\alpha\Omega$ dynamo in a spherical shell with sharp gradients
of the turbulent magnetic diffusivity and the strength of the $\alpha$ effect
at the bottom of the convection zone; for details we refer to \citet{jouetal08}. The potential functions $A$ and $B$ are expanded according
to Eqs. (\ref{eq:A-dec})--(\ref{b_n})  in Sect. \ref{sec_model},
 and the integration
domain is now radially bounded by  $x_{\mathrm{b}}=0.65$ at the bottom and $x_{\mathrm{t}}=1$ at the top.

\begin{figure*}
\begin{centering}
\includegraphics[width=0.8\textwidth]{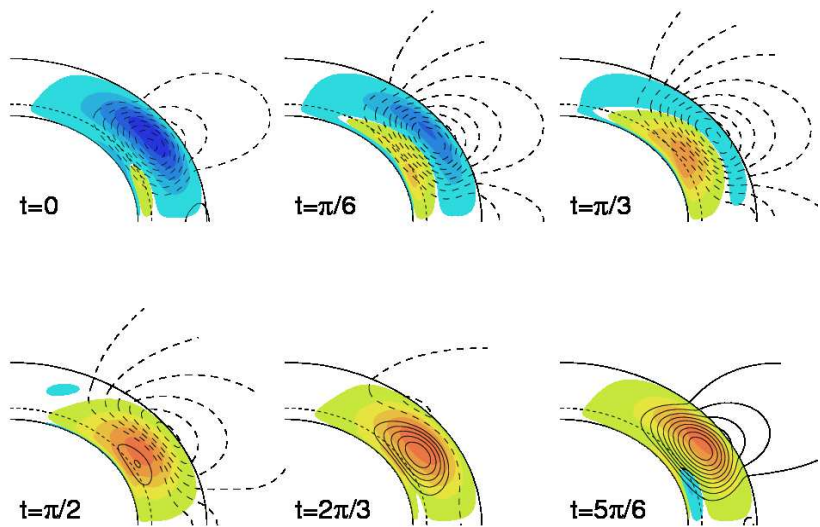}
\par\end{centering}
\caption{\label{snapshots:test_case_B}
As Fig.~\ref{fig1} (top and middle), but for test case B of \citet{jouetal08}
at the critical $\alpha$-effect dynamo number with a spectral
resolution of $16\times16$ modes.
}
\end{figure*}

In \citet{jouetal08}, the strength of the $\alpha$ effect is regulated by a
dynamo number, $C_\alpha$. The different codes are compared by indicating in tables
the critical $\alpha$-effect dynamo number, $C_{\alpha}^{\mathrm{crit}}$, at which exponentially
growing solutions appear, and the corresponding oscillation frequency,
$\omega=2\pi/T$.
In addition, butterfly diagrams and
the evolution of the fields in the meridional plane are shown.
Our values of $C_{\alpha}^{\mathrm{crit}}$ and $\omega$ for different spectral resolutions are given in Table \ref{table:test_case-B},
\begin{table}
\caption{\label{table:test_case-B}
Test case B of \citet{jouetal08} $^{\mathrm{a}}$
}
\begin{centering}
\begin{tabular}{ccc}
\hline 
Resolution & $C_{\alpha}^{\mathrm{crit}}$ & $\omega$\tabularnewline
\hline
$8\times8$ & 0.443 & 180.5\tabularnewline
$10\times10$ & 0.4175 & 175.1\tabularnewline
$12\times12$ & 0.4095 & 172.2\tabularnewline
$13\times13$ & 0.411 & 172.4\tabularnewline
$14\times14$ & 0.4122 & 172.7\tabularnewline
$16\times16$ & 0.4125 & 172.9\tabularnewline
\hline
\end{tabular}
\par\end{centering}
\begin{list}{}{}
\item[$^{\mathrm{a}}$]
Critical values of the dynamo number, $C_{\alpha}^{\mathrm{crit}}$, and the oscillation
frequency at the dynamo onset, $\omega$, for different radial and latitudinal spectral
resolutions. $\omega$ is measured in units of the inverse magnetic diffusion time.
\end{list}
\end{table}
and Fig.~\ref{snapshots:test_case_B}
shows the temporal evolution of the toroidal and poloidal parts of the field
(i.e., of the unstable eigenmode) at the critical dynamo number.
The values in Table \ref{table:test_case-B} are in best agreement with those
given in the corresponding table, Table 3, of \citet{jouetal08}.
Similarly, the evolution shown in Fig.~\ref{snapshots:test_case_B} is apparently identical
to that shown in the corresponding figure, Fig.~7, of \citet{jouetal08};
the same applies to the simulated butterfly diagrams (not shown here).

{
\subsection{Convergence of crititical dynamo numbers and dynamo
periods for the models of Sects.~\ref{sec_delta-Omega-dynamo}
and \ref{sec_delta-W-dynamo}}}

\begin{figure*}
\begin{centering}
\includegraphics[angle=-90,width=0.4\textwidth]{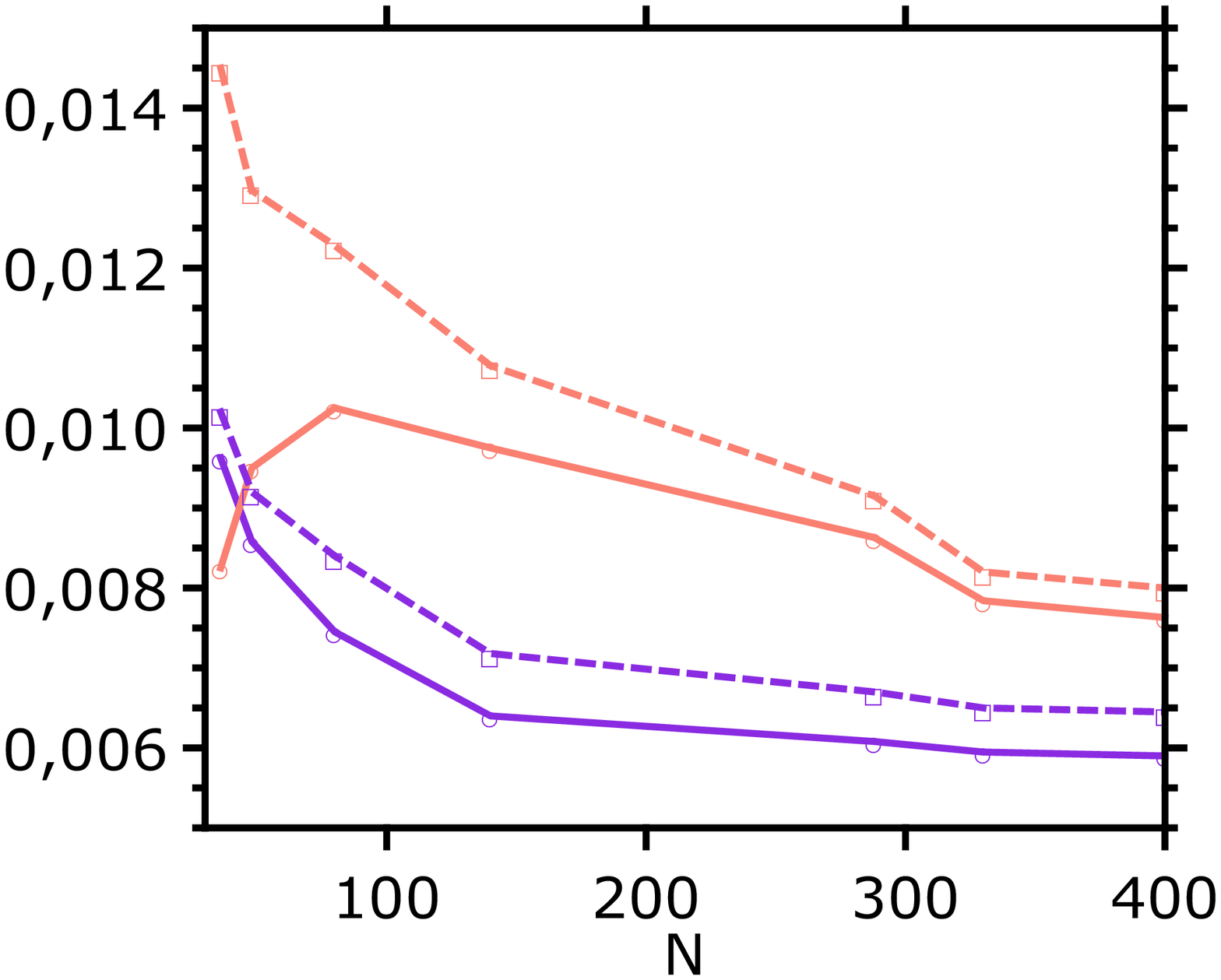}
\hspace{1em}
\includegraphics[angle=-90,width=0.4\textwidth]{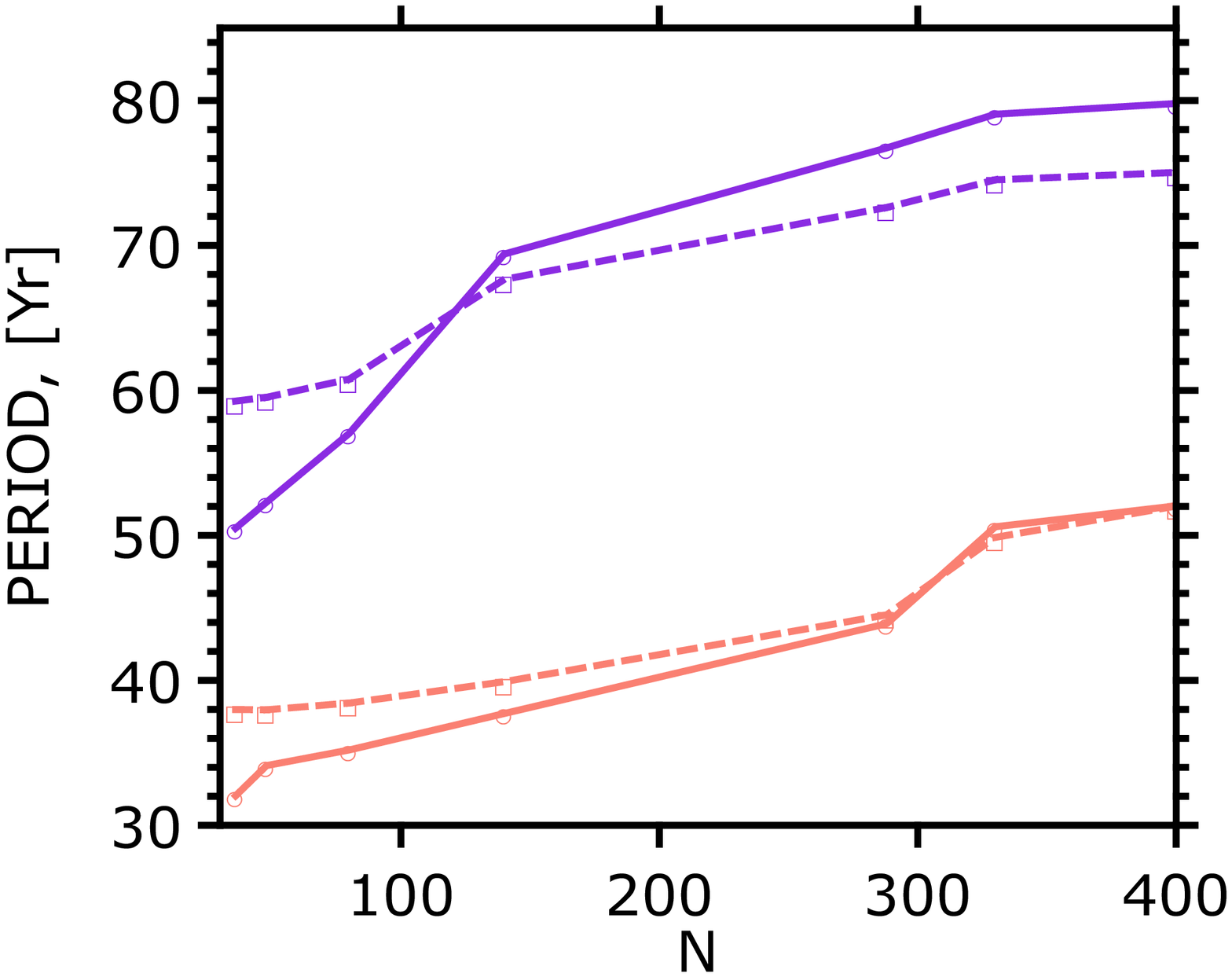}
\caption{\label{fig:conv13}{ Convergence of the critical dynamo numbers {\em (left)}
and associated dynamo periods {\em (right)} for the models of Sects.~\ref{sec_delta-Omega-dynamo} and \ref{sec_delta-W-dynamo}. $N$ is the total number of modes taken into account.
Calculations were done for resolutions of $6\times 6$, $6\times 8$, $8\times 10$, $10\times14$, $16\times 18$,
$15\times 22$, and $14\times 30$ modes in the radial and latitudinal bases, respectively
(in addition, the dipolar symmetry was taken into account, so that the highest latitudinal resolution
is actually 50).
Solid lines refer to
the $\delta^{(\Omega)}\Omega$ dynamo model and dashed lines to the
$\delta^{(W)}\Omega$ dynamo model, and blue color corresponds to $u_0=15\,\mathrm{m/s}$ and red color to $u_0=25\,\mathrm{m/s}$.}
}
\end{centering}
\end{figure*}

{Fig.~\ref{fig:conv13}}
{shows the convergence of the critical dynamo numbers
(where the first dipolar mode becomes unstable) and associated dynamo periods
for the $\delta^{(\Omega)}\Omega$ dynamo model considered in Sect.~\ref{sec_delta-Omega-dynamo}
and for the  $\delta^{(W)}\Omega$ dynamo model considered in Sect.~\ref{sec_delta-W-dynamo}.
The amplitude of the meridional flow is $u_0=15\,\mathrm{m/s}$ and $u_0=25\,\mathrm{m/s}$;
$u_0=15\,\mathrm{m/s}$ is the value we used most, and
 $u_0=25\,\mathrm{m/s}$ is the highest meridional-flow amplitude that we considered, corresponding
to the largest magnetic Reynolds number in the study. High Reynolds numbers are known to cause numerical problems.}

\section{Definitions of the functions $f_i^{(a)}$ and $f_i^{(d)}$}
\label{appendix_definitions}

 Here we give the definitions of the functions $f_i^{(a)}$ and $f_i^{(d)}$ that are used in the
representation
of the turbulent electromotive force $\vec{\mathcal{E}}$. For details of the calculations
we refer to \citet{pip08}.
\begin{eqnarray*}
f_{1}^{(a)} & = & \frac{1}
{4\Omega^{*\,2}}\left[\left(\Omega^{*\,2}+3\right)\frac{\arctan\Omega^{*}}
{\Omega^{*}}-3\right]\,,\\
f_{3}^{(a)} & = & \frac{1}
{4\Omega^{*\,2}}\left[\left(\left(\varepsilon-1\right)\Omega^{*\,2}+\varepsilon-3\right)\frac{\arctan\Omega^{*}}
{\Omega^{*}}+3-\varepsilon\right]\,,\\
f_{2}^{(d)} & = & \frac{1}
{4\Omega^{*\,2}}\left[\left(\left(\varepsilon-1\right)\Omega^{*\,2}+3\varepsilon+1\right)\frac{\arctan\Omega^{*}}
{\Omega^{*}}\right.\\
&&-\left(3\varepsilon+1\right)\Bigg]\,,\\
f_{4}^{(d)} & = & \frac{1}
{2\Omega^{*\,3}}\left[\left(2\Omega^{*\,2}+3\right)-3\left(\Omega^{*\,2}+1\right)\frac{\arctan\Omega^{*}}
{\Omega^{*}}\right]\,.
\end{eqnarray*}

\end{appendix}

\end{document}